\documentclass[12pt]{article}
\input epsf.tex
\usepackage{epsfig,psfrag}

\catcode`@=12
     
\def\nd{\nu \phantom{} DE}
\def\aa{{\cal A}}
\begin{document}
\thispagestyle{empty}
\begin{flushright}
DO-TH 06/06\\
July 26, 2006\
\end{flushright}
\vspace{0.5in}
\begin{center}
{\large \bf Neutrino Phenomenology, Dark Energy and Leptogenesis
from \\ pseudo-Nambu-Goldstone Bosons
\\}
\vspace{1.0in}
{\bf C.T. Hill${}^{1}$, I. Mocioiu${}^{2}$, E.A. Paschos${}^{3}$ and
U. Sarkar${}^{3,4}$ \\}
\vspace{0.2in}
{\sl ${}^{1}$ Fermi National Laboratory,  P.O. Box 500, Batavia,
Illinois 60510, USA\\}
{\sl ${}^{2}$ Pennsylvania State University, 104 Davey Lab,
University Park, PA 16802-6300, USA\\}
{\sl ${}^{3}$ Institut f\"ur Physik, Universit\"at Dortmund,
D-44221 Dortmund, Germany\\}
{\sl ${}^{4}$ Physical Research Laboratory, Ahmedabad 380009, India\\}
\vspace{.5in}
\end{center}

\begin{abstract}
We consider a model of dynamical neutrino masses 
via the see-saw mechanism. Nambu-Goldstone bosons (majorons) arise associated
with the formation of the heavy right-handed majorana masses.
These bosons then acquire naturally soft masses (become pNGB's) 
at loop level via the Higgs-Yukawa
mass terms. These models, like the original neutrino pNGB quintessence
schemes of the 1980's \cite{4,frieman}
that proceed through the Dirac masses,  are natural, 
have cosmological implications through mass varying neutrinos, 
long range forces, and provide a  soft
potential for dark energy. We further argue that these models 
can explain leptogenesis 
naturally through the decays of the right-handed neutrinos.

\end{abstract}

\newpage
\baselineskip 22pt
\section{Introduction}

The conventional explanation for neutrino mass is the
see-saw mechanism \cite{seesaw}.
In this scheme, sterile right-handed neutrinos have very large
(grand unification scale) majorana
masses $M_i$. The standard model is then a low energy
effective theory, obtained by integrating out
the right-handed neutrinos, leading to suppressed induced
Majorana masses for the left-handed neutrinos, of order
$v^2/M$, where $v$ is the electroweak scale VEV, $v=175$ GeV
or the typical Dirac mass for leptons.
The large $M_i$ break the global lepton numbers
of the right-handed neutrinos. It is natural
to hypothesize that this symmetry breaking is dynamical, much like the 
chiral dynamics of QCD, and is quite similar to
the Cooper pairing in a superconductor.
The spontaneous breaking of these
global symmetries will produce massless 
Nambu-Goldstone bosons (NGB's). In the present case such NGB's associated
with majorana mass generation are
termed ``majorons" \cite{gelmini}.
It is possible that a whole category of new particles, NGB's
such as 
axions \cite{1}, majorons, familons, etc., 
exists in nature, reflecting the generic breaking of
multitudinous global symmetries.  

The spontaneously broken global symmetries of the right-handed
neutrinos will generally be explicitly broken as well. This explicit breaking
must, at least, arise in the Yukawa couplings of the right-handed 
and left-handed neutrinos to the Higgs field of the standard model, i.e.,
in the Dirac mass terms that marry the left-handed
$I=1/2$ and right-handed sterile neutrinos. 
In the present paper we will assume that this is the primary source of the
explicit global symmetry breaking, and that any other sources of such breaking
do not lead to significantly larger effects than these.
It then follows that the majorons will become 
pseudo-Nambu-Goldstone bosons (pNGB's) { at least} via loop diagrams that
involve insertions of the various Yukawa vertices in the presence of the Higgs
vacuum expectation value (VEV).  These loops require sufficiently many
insertions that they become finite, and the pNGB masses 
are then ``calculable,'' in a sense discussed by Georgi and Pais \cite{georgi}.
This was further developed by Hill and Ross 
\cite{hillross} to construct technically
natural models, involving 
CP-violation and other symmetry breaking effects, that give rise to 
calculable novel
long range forces in the context of the Standard Model.
The finiteness of the induced pNGB masses is analogous to 
what happens in deconstructed extra dimensions \cite{decon}.
E.g., for QED in $D=5$ an NGB arises that is  $\int dx^5 A_5$, the 
Wilson line of the fifth component 
of the photon vector potential integrated over the compact fifth dimension.
In deconstruction, this is finite when $N\geq 3$ lattice slices are taken
for the fifth dimension.  Extra dimensions or
deconstruction  offer a method of solving the naturalness
problems associated with large decay constants of axions and
other pNGB's \cite{h2}.

In the case of majoronic pNGB's  the mass scales of majorons are
generally small and majorons will have { large finite} 
Compton wavelengths.  
Neutrino majorons can thus have astrophysical/cosmological implications.
This was first recognized many years ago as a general
mechanism to induce large distance or late-time effects
in the { not-so early universe by Hill, Schramm and Fry \cite{4},
 and was subsequently
developed in greater detail by Frieman, Stebbins, Waga, Kolb et.al.
\cite{frieman}}.   These models, which exploit
spontaneous chiral symmetry breaking associated with Dirac masses,
have the virtue of naturalness, and also, unlike dilatonic schemes
and a host of random scalar potential models, 
we {\em have prior experience} in
physics with pNGB's (eg., the pion is a pNGB).
These models are amongst the original, 
and are well motivated as they are the first natural, 
quintessence models \cite{4,frieman,5}.  They are also the original 
models of mass varying neutrinos (``MAVAN's''). This general phenomenon
has been largely rediscovered in recent years 
\cite{mavans,kap,lg}, and the original models can provide
a natural origin to an ``acceleron'' field. The present analysis
incorporates many features of \cite{4,frieman} 
within the context of the seesaw mechanism
and the dynamical generation of right-handed majorana masses. 

We further investigate the possibility of a majoronic pNGB 
as an ``acceleron,'' studying its implications for the leptonic sector,
i.e., leptogenesis.

\section{Two Generation pNGB Model}

Consider an extension of the standard model with two right-handed
neutrinos $N_1$ and $N_2$, which are singlets under the standard model
gauge groups. There are no majorana masses for these fields and hence
lepton number is not explicitly broken at this level. 
We introduce two singlet
scalar fields, $\Phi_1(x)$ and $\Phi_2(x)$, which are also singlets
under the standard model gauge groups. We postulate that
these fields couple with the
right-handed neutrinos as:
\begin{equation}
{\cal L}_M = { 1 \over 2} \alpha_1 \bar N_1 N_1^c \Phi_1 +
{ 1 \over 2} \alpha_2 \bar N_2 N_2^c \Phi_2 .
\end{equation}
Then these scalars will acquire vacuum expectation values (VEV's)
and will give Majorana masses to the right-handed neutrinos.
The model possesses a global $U(1)_1 \times U(1)_2$ symmetry.
Under $U(1)_1 \times U(1)_2$, the quantum numbers of the fields are
$N_1 \equiv (1,0)$, $N_2 \equiv (0,1)$, $\Phi_1 \equiv (2,0)$ and
$\Phi_2 \equiv (0,2)$ respectively. 
When the $\Phi_i(x)$ acquire VEV's, these global
symmetries are broken and there will be two Nambu-Goldstone bosons.

We assume that the fields $\Phi_i$ acquire
VEV's, and it is then
useful to parametrize them
as nonlinear $\sigma$-model fields in terms of the NGB's, $\phi_i$
and decay constants, $f_i$,  
\begin{equation}
\langle \Phi_i \rangle = (f_i/2\sqrt{2})\exp(2i\phi_i/f_i)
\end{equation}
and we have
$ \alpha _i \Phi_i \to  M_i e^{2 i \phi_i/f_i} $ .
Here we assume a common value for
the decay constants  $f_i = f \sim M_i$. The $\phi_i$
are massless NGB's at this stage.

The $U(1)_1 \times U(1)_2$ symmetry  will generally be explicitly broken
by the Yukawa interactions of the right-handed neutrinos with the left-handed
leptons ($\ell_i \equiv \{ \nu_i, e^-_i \}$)
through the usual standard model Higgs doublet ($H_0$). 
{ If we allow such symmetry breaking terms, there will be
divergent contributions to the mass of the Nambu-Goldstone bosons.
Only when the original symmetry is broken by soft terms,
which may originate from electroweak symmetry
breaking, then the mass of the pNGB remains finite and small. For this reason
we introduce two new Higgs doublets $H_1$ and $H_2$, transforming
under the $U(1)_1 \times U(1)_2$ as $(+1,-1)$ and
$(-1,+1)$ respectively, to make
the theory invariant under $U(1)_1 \times U(1)_2$. Assigning
the $U(1)_1 \times U(1)_2$
quantum numbers $(1,0)$ for $\ell_1 \equiv \pmatrix{ \nu_e \cr e^-}$
and $(0,1)$ for $\ell_2 \equiv \pmatrix{ \nu_\mu \cr \mu^-}$
the $U(1)_1 \times U(1)_2$ invariant Yukawa interactions are
given by
\begin{eqnarray}
{\cal L}_{mass} &=& f_{11} \bar N_1 \ell_1 H_0 + f_{12} \bar N_1
\ell_2 H_1
+f_{21} \bar N_2 \ell_1 H_2 + f_{22} \bar N_2 \ell_2 H_0 .
\end{eqnarray}
The soft terms may also originate from some new physics at
higher energies.} The Higgs scalars $H_1$ and $H_2$ are 
candidates for 
cold dark matter of the universe, since they cannot
decay into quarks or light leptons \cite{barhall}.

Once the Higgs fields develop electroweak scale VEVs these
become Dirac mass terms. The complete neutrino mass matrix is then:
\begin{eqnarray}
-{\cal L}_{mass} &=& { 1 \over 2} M_1 \bar N_1 N_1^c e^{2 i \phi_1 /f} +
{ 1 \over 2} M_2 \bar N_2 N_2^c e^{2 i \phi_2/f}
+ m e^{i \alpha} \bar N_1 \nu_1 + m \epsilon e^{i \beta} \bar N_1
\nu_2 \nonumber \\
&&+ \lambda m \epsilon^\prime e^{i \gamma} \bar N_2 \nu_1
+ \lambda m e^{ i \xi} \bar N_2 \nu_2 .
\label{dirac}
\end{eqnarray}
We've introduced an overall
Dirac mass parameter $m$ and scaling parameters 
$\lambda, \epsilon, \epsilon^\prime$, and
we've also included all the phases which contribute to CP violation.

{ A three Higgs doublet model like the one presented above could
also have interesting collider phenomenology, as it has a much richer
electroweak symmetry breaking structure than the Standard Model. For
our purposes it is sufficient to have the three scalars develop an
electroweak scale vev which would lead to the Dirac masses above.}

Note that we can make phase redefinitions
on the neutrino fields $N_i$ and $\nu_i$ to try to absorb  
the majoron fields out of the mass matrix. 
Transforming the neutrinos along with the
redefinition of the CP phases, we can bring the
mass matrix into the form:
\begin{eqnarray}
-{\cal L}_\mu &=& {1 \over 2} M_1 \bar N_1 N^c_1 e^{2 i \phi/f}+
{1 \over 2} M_2 \bar N_2 N^c_2 +
m \bar N_1 \nu_1 + m \epsilon e^{i \eta  } \bar N_1
\nu_2 \nonumber \\
&&+ \lambda m \epsilon^\prime e^{i \eta } \bar N_2 \nu_1
+ \lambda m \bar N_2 \nu_2 + h.c.
\label{full}
\end{eqnarray}
where $2 \eta = \gamma - \alpha + \beta - \xi$
and $\phi = \phi_1 -\phi_2$.  It
is not possible by further transformations to remove the field
$\phi$ or phase $\eta$ from the mass terms.  
The combination $\phi_1+\phi_2$ has been absorbed out
of the mass terms and 
remains a massless NGB, while the field $\phi =
\phi_1 - \phi_2$ becomes a pNGB, due to the explicit breaking
of the $U(1)_{1 - 2}$ symmetry by the Higgs-Yukawa terms.

\begin{figure}[!h]
\begin{center}
{
\psfrag{a}{$\nu_2$}
\psfrag{b}{$\nu_1$}
\psfrag{c}{$N_1$}
\psfrag{d}{$N_2$}
\psfrag{x}{$m\epsilon e^{i\eta}$}
\psfrag{y}{$m$}
\psfrag{z}{$\lambda m$}
\psfrag{t}{$\lambda m \epsilon' e^{-i\eta}$}
\psfrag{m}{$M_1/f^2$}
\psfrag{n}{$M_2$}
\psfrag{u}{$\phi$}
\epsfig{file=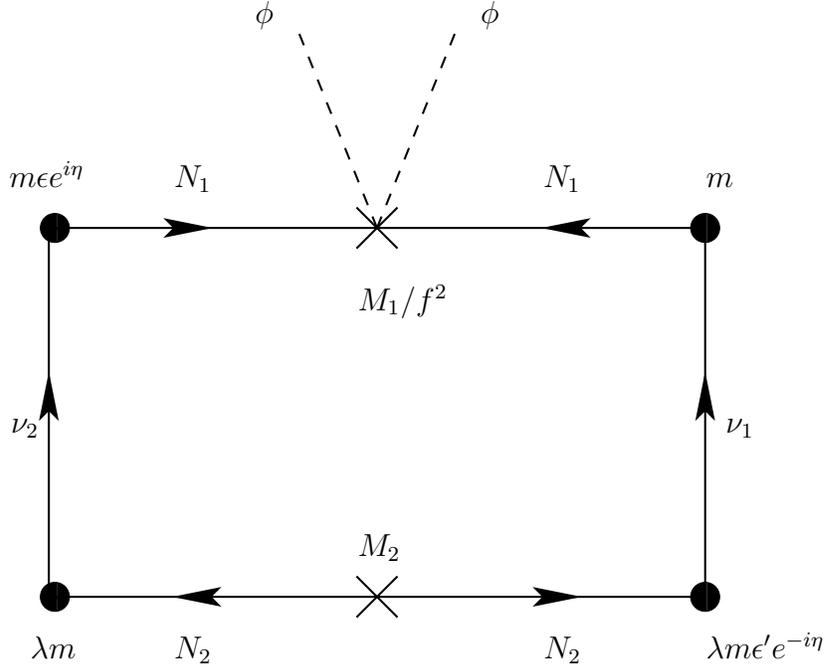,width=10cm}
}
\caption{Loop diagram for the effective potential
}
\label{figure1}
\end{center}
\end{figure}

The interaction of the scalar field through the Majorana and Dirac terms
generates a Colemann-Weinberg effective potential for $\phi$.
This can be estimated through the
leading loop in Fig. 1. It has the remarkable property that the symmetry
structure of the theory makes the loop finite. The reason is that the $\phi$
field could be eliminated if any of the vertices is set to zero. 
{ The vertices in this diagram come after electroweak
symmetry breaking. Before the electroweak symmetry breaking
the Nambu-Goldstone boson corresponding to $U(1)_{1-2}$ remains
massless. After the electroweak symmetry breaking these vertices
appear as soft terms and hence they cannot introduce 
non-renormalizable terms that need to be counter balanced.}

{
Note that the phase $\eta$ disappears from the potential and the field $\phi$ cannot be
removed from the diagram by rephasing of any fields. More generally, 
the diagram is invariant under any field rephasings involving $\phi_i$ or other CP phases. }

An explicit calculation gives: 
\begin{equation}
V_{eff} (\phi^2) = - {m^4 \lambda^2 \epsilon \epsilon^\prime \over
2  \pi^2} {M_1 M_2 \log \left( {M_1^2 \over M_2^2} \right) 
\over M_1^2 - M_2^2 } \cos \left({ 2 \phi \over f} \right).
\end{equation}
The potential for the two generation model depends on all coupling constants,
is weakly dependent on the heavy scales $M_i$, 
 and has minima at $\phi = 0, \pi f,
2 \pi f, \cdot \cdot \cdot$. We can expand the potential around one minimum
obtaining a constant term, a mass term and higher orders in $\phi$
giving interactions:
\begin{equation}
V_{eff} (\phi) = {m^4 \lambda^2 \epsilon \epsilon^\prime \over
2 \pi^2} {M_1 M_2 \log \left( {M_1^2 \over M_2^2} \right) 
\over M_1^2 - M_2^2 } 
\left( -1 + 2 {\phi^2 \over f^2} - O({\phi^4 \over f^4}) +
\cdot \cdot \cdot \right).
\end{equation}
Thus the induced mass of the field $\phi$ is now:
\begin{equation}
m_\phi = {m^2 \lambda \sqrt{\epsilon \epsilon^\prime} \over
\pi f} {M_1 M_2 \log \left( {M_1^2 \over M_2^2} \right) 
\over M_1^2 - M_2^2 } .
\end{equation}
{ If $M_1=M_2$ all the $M_i$ dependent factor in the equation above becomes 1. }
As mentioned earlier, the symmetry at the scale $f \sim M_i$ protects
the mass of the pseudo Nambu-Goldstone boson, so an explicit breaking
of the symmetry at the scale $m$ can generate a mass of the order of
$m^2/f$. Thus the mass of the scalar is very small. 
{ We now turn to the see-saw mechanism, 
long-range force, leptogenesis and the origin of dark energy
in this model}.

\section{Neutrino Masses and Long Range Forces Among Neutrinos}

We shall next study the neutrino masses in this scenario. If we
ignore the small mass varying effect on the neutrino masses coming
from the pNGB, the time-development of the light states is
determined by the matrix
\begin{equation}
- {\cal L}_{eff} = \overline{\nu^c} m_D^T M_R^{-1} m_D\nu = {m^2 \over M}\left[ {\pmatrix{\overline{\nu^c_1} & \overline{\nu^c_2}} \atop }
\pmatrix{ 1 + (\lambda \epsilon^\prime)^2
e^{2 i \eta } & e^{ i \eta} ( \epsilon + \lambda^2 \epsilon^\prime) \cr
e^{ i \eta} ( \epsilon + \lambda^2 \epsilon^\prime) & \lambda^2 + \epsilon^2
e^{2 i \eta }  } \pmatrix{\nu_1 \cr \nu_2} \right] 
\end{equation}
{ where we have assumed $M_1\simeq M_2=M$.
This mass matrix can be diagonalized to
\begin{equation}
m^{diag} = \pmatrix{m_1 e^{ i \theta_1} & 0 \cr 0 &m_2 e^{ i
\theta_2}} \,.
\end{equation}
The mixing angle is:
\begin{equation}
\tan 2 \theta =-
\frac{2(\epsilon+\epsilon'\lambda^2)\sqrt{(1+\epsilon^2)^2+
(1+\epsilon'^2)^2\lambda^4+2 (1+\epsilon^2)(1+\epsilon'^2)\lambda^2
\cos2\eta}}{1-\epsilon^4-\lambda^4(1-\epsilon'^4)-2
(\epsilon^2-\epsilon'^2)\lambda^2\cos2\eta}\, .
\end{equation}

We would like to reproduce the maximal mixing observed for atmospheric
neutrinos. One possibility that immediately gives this result is to
have $\epsilon,\epsilon' \ll 1$ and $\lambda \approx 1$, up to terms
of order $\epsilon_i^2$.
Keeping only the leading terms in $\epsilon$, $\epsilon'$ or $1-\lambda$, the
mixing angle becomes:
\begin{equation}
\tan 2 \theta =
\frac{2(\epsilon+\epsilon')\cos\eta}{-2 (1-\lambda) + (\epsilon^2-\epsilon'^2)\cos2\eta}\, .
\end{equation}
The neutrino masses are given by:
\begin{equation}
m_1\simeq {m^2 \over M} \left[ 1+(\epsilon+\epsilon')\cos\eta \right]
\end{equation}
and
\begin{equation}
m_2\simeq {m^2 \over M} \left[ 1-(\epsilon+\epsilon')\cos\eta \right]\,.
\end{equation}

The mass squared difference, which
should be in the atmospheric range of $\sim 2\times 10^{-3} {\rm
eV}^2$, is of order $\epsilon_i$: $\Delta m^2=-4 ~m^4
(\epsilon+\epsilon')\cos\eta /M^2$. Since the $\epsilon_i$ parameters were
assumed to be small, the required right-handed neutrino mass scale
should be lower than in standard see-saw type scenarios.
}

As we go a step further and keep the linear term in the scalar field, we
obtain in addition an interaction between the neutrinos and scalar field.
The exchange of the scalar particle between the eigenstates results in a new force whose
range is at our disposal. 
We are free to vary the mass of the scalar by
selecting values for the decay constant $f$. The final mass and interaction
Lagrangian,{ up to terms linear in $\epsilon,\epsilon'$}, is given by
\begin{equation}
{\cal L} = {m^2 \over 2 M} ~ \overline{\Psi^c} \left\{ \pmatrix{m_1
&0\cr 0 & m_2} 
- i {\phi \over f} \pmatrix{e^{i\alpha_1} &-e^{i\alpha_2}
\cr -e^{i\alpha_2}&e^{i\alpha_1} } \right\} \Psi + H.C.
\end{equation}
The { off-diagonal} long range force predicted by the couplings of the pNGB in this
model could have direct consequences in neutrino oscillation experiments
\cite{bhm05} or in cosmology \cite{4,frieman}. 

\section{Lepton Asymmetry of the Universe}

\begin{figure}[!t]
\begin{center}
\vskip .5in
\epsfxsize14cm\epsffile{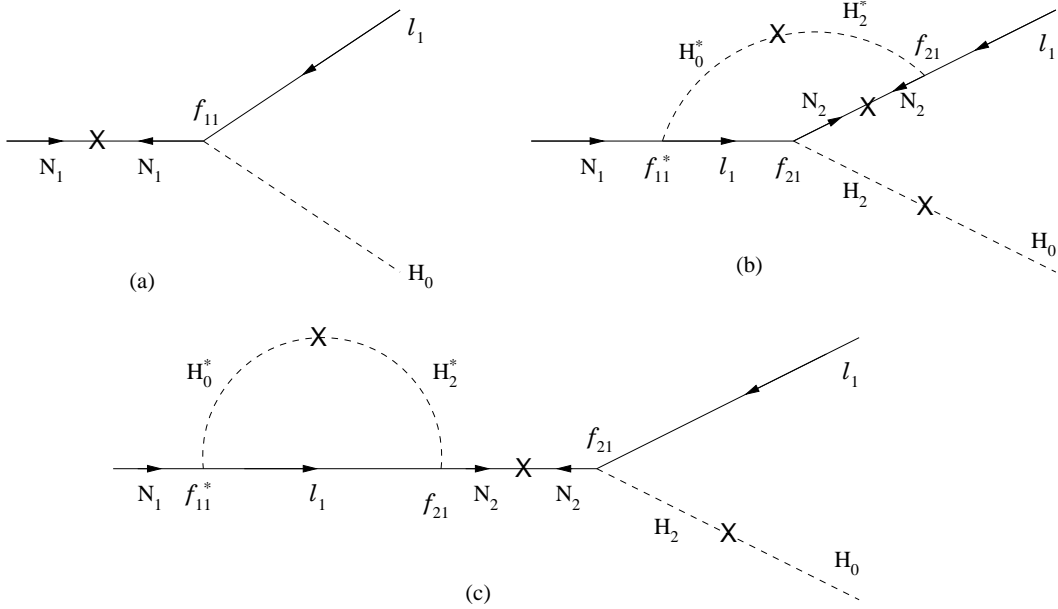}
\caption{Tree level and one-loop diagrams that contributes
to the generation of lepton asymmetry by the decays of $N_1$.
}
\label{figure2}
\end{center}
\end{figure}

The decays of the right-handed neutrinos
\begin{eqnarray} 
N_i &\to& \ell_j + H_a^\dagger \nonumber \\
&\to& \ell_j^c + H_a  \nonumber
\end{eqnarray}
violate lepton number and CP violation comes from the CP phase $\eta$.
There are vertex-type \cite{lepto,lepto1} and self-energy \cite{lepto2} 
one-loop diagrams that interfere with the tree-level decays to
give an asymmetry as shown in figure \ref{figure2}. 
The generation of a baryon asymmetry requires
one more important ingredient, namely, the mixing of the Higgs
doublets $H_{0,1,2}$. The quartic couplings $|H_0|^2 |H_i|^2$
give all the required couplings which are not suppressed. We
denote the mixing of $H_0$ with $H_i$ by $V_{0i}$. 

We assume that $N_2$ is heavier than $N_1$ ($M_1 < M_2$)
so that $N_2$ decays first and at a
later time the decays of $N_1$ contribute to the asymmetry. The interference
of the tree-level and one-loop diagrams in figure \ref{figure2}
generate the asymmetry. The cross on a fermion line
indicates a Majorana mass insertion and the cross on a Higgs line 
refers to the mixing between
the scalars. The tree-level and one-loop
diagrams that contribute to the
generation of lepton asymmetry in the decays of $N_1$ are
shown in figure \ref{figure2}. There will be similar diagrams
contributing to the process $N_1 \to \ell_2 + H_1$.

The interference between the diagrams give the asymmetry

\vbox{
\begin{eqnarray}
\delta &=&  { \Gamma ( N_1 \to \ell H^\dagger) -
\Gamma ( N_1 \to \ell^c H) \over \Gamma ( N_1 \to \ell H^\dagger) +
\Gamma ( N_1 \to \ell^c H)} \nonumber \\
&=&{3 \over 16 \pi} {M_1 \over M_2} ~{m^2 \lambda^2 \over
v_1^2 + v_0^2 \epsilon^2 } ~\left[ |V_{02}|^2 {v_1^2 \over
v_2^2} {\epsilon^\prime}^2 - |V_{01}|^2 \epsilon^2
\right]~ \sin 2 \eta.
\end{eqnarray}}
\noindent where $v_a = \langle H_a \rangle, a = 0,1,2$.
{ One interesting feature of this model is that the
CP violating phase $\eta$ is solely responsible for the generation
of the lepton asymmetry of the universe and the sign of the asymmetry
is also determined in terms of the parameters of this model. 
This phase may also be observed in lepton number violating
processes like neutrinoless double beta decay.}

At temperatures above $T > 10^9$ GeV, 
the out-of-equilibrium condition can be satisfied
\begin{equation}
\Gamma_{N_1} = {|f_{1 i}|^2 \over 16 \pi} M_1 < H(M_1) =
1.7 \sqrt{g_*} {T^2 \over M_{Pl}} \hskip .5in {\rm at}~T = M_1 .
\end{equation}
However, for thermal production of the right-handed neutrinos,
the interaction rate ($\Gamma_{N_1}$) should not be much smaller than the
expansion rate of the universe ($H$). Considering 
the requirement of thermal production and subsequent wash out of the 
asymmetry, in addition to the suppression factor 
$\kappa = \Gamma_{N_1}/H(M_1)$,
which enters if the decay rate is slightly greater than the
expansion rate of the universe,
the amount of $(B-L)$ asymmetry is given by
\begin{equation}
{n_{B-L} \over s} \approx - {40 \over g_*~ \pi^4}~{\delta 
\over \kappa }
\end{equation}
The sphaleron interactions now convert this ${(B-L)}$ asymmetry to a baryon
asymmetry 
\begin{equation}
{n_B \over s} = {24 + 4 n_H \over 66 + 13 n_H} {n_{B-L} \over s},
\end{equation}
where $n_H$ is the number of Higgs doublets in the standard
model. This can then generate the required amount of baryon asymmetry
of the universe.
 
\section{Origin of Dark Energy}

The basic idea behind the neutrino dark energy ($\nd$) models is that
the neutrino mass varies as a function of a light scalar field 
$\aa$ (the acceleron) \cite{mavans} and the mass 
of the acceleron today should be less than $\sim (10^{-4} ~ {\rm eV})$.
In the present model, the pNGB $\aa = (i \phi)$
corresponding to the soft $U(1)_{A-B}$
global symmetry breaking takes the role of the acceleron ($\aa$). Thus the
mass matrix becomes
\begin{eqnarray}
{\cal L}_\mu &=& {1 \over 2} M_1(\aa) \bar{N_1^c} N_1 +
{1 \over 2} M_2 \bar{N_2^c} N_2 +
m \bar N_1 \nu_1 + m \epsilon \bar N_1
\nu_2 \nonumber \\&&
+ \lambda m \epsilon^\prime \bar N_2 \nu_1
+ \lambda m \bar N_2 \nu_2 + h.c.
\label{full1}
\end{eqnarray}
where $M_1(\aa)$ is the mass of the heavy
right-handed neutrino $N_1$ that varies explicitly
with the acceleron field $\aa$.
It is interesting to note that $M_1(\aa)$ is specified and
the effective neutrino mass also varies explicitly 
$m_\nu (\aa) = m^T M^{-1} (\aa) m$, 
as required by some models of mass varying neutrinos. The present
model has the same generic problems like any other models of
mass varying neutrinos \cite{peccei}, some of which have been
taken care of in variants of this model \cite{models}, 
but the main feature of
the present model is that it explains the origin of the acceleron field
as the pseudo Nambu-Goldstone boson (pNGB).

It is also
possible to construct scenarios where the pseudo Nambu-Goldstone boson
acts as cold dark matter \cite{Das:2006ht} or can lead to
non-standard structure-formation due to the strong long range
interaction among neutrinos. 

\section{Summary}

{ A large Majorana scale for neutrino masses is a necessary 
component of the see-saw mechanism.  At this very high energy 
scale there could be global symmetries which are broken in order to 
produce Majorana masses.  Remnants of the symmetries may exist 
at low energy.  In this article we proposed a 
generic model with global symmetry. The 
breaking of this global symmetry gives masses to right-handed 
neutrinos and produces Nambu-Goldstone bosons. 

In this framework we investigated a model with two generations 
of neutrinos and scalar particles.  We showed that the 
Dirac masses of the neutrinos appearing from the electroweak 
symmetry breaking implies soft breaking of this global symmetry,
which generates an effective potential that 
is finite.  The Nambu-Goldstone boson then acquires a mass and 
produces a long-range force between neutrinos.  
Their masses have an 
explicit dependence on the scalar field, which may bring 
additional density effects producing cosmological consequences.

The model has other attractive properties.  The decays of 
the heavy neutrinos generate a lepton asymmetry consistent 
with the scenario of leptogenesis.  It also describes the 
masses and mixings of light neutrinos.  The
pNGB can play the role of the ``acceleron''
field introduced in models of mass varying neutrinos. 
}

\noindent {\bf Acknowledgement}

\noindent 
C.T. Hill thanks Dortmund University for the award of a Gambrinus
Fellowship and its hospitality during the early stages of this work. 
E.A. Paschos thanks the Theory Group of Fermilab for its hospitality.
U. Sarkar thanks the Institut f\"ur Physik, Univ. Dortmund and
Alexander von Humboldt Foundation for their support. The work of I. 
Mocioiu was supported in part by the National Science Foundation 
grant PHY-0555368 adn the work for two of us (EAP
and US) by BMBF, Bonn under contract 05 HT 4PEA/9.

\newpage
\bibliographystyle{unsrt}

\end{document}